# Empirical Research of a Novel Weighted Network: Knowledge Network

X.F. Jiang*
*Library, Nanjing University, 210093, P.R. China

Abstract: Networked structure is included in a wide range of fields such as biological systems, World Wide Web and technological infrastructure. A deep insight into the topological complexity of these networks has been gained. Some works start to pay attention to the weighted network, like the world-wide airport network and the collaboration network, where links are not binary, but intensive. Here, we construct a novel network named knowledge network, through which we take the first step to uncover the topological structure of the knowledge system. Furthermore, the network is extended to the weighted one by assigning weights to the edges. These results provide a framework to understand the hierarchies and organizational principles in knowledge system, and the interaction between the intensity of edges and topological structure. This system also provides a good paradigm to study weighted networks.

Many systems of current interest to the scientific community can usefully be represented as networks (*1,2,3*), consisting of a set of nodes or vertices representing. Instances include the Internet (*4*), social networks (*5*), food webs (*6*) and human disease network (*7*). Specially, one of interesting fields of complex networks is specifically focused on the information networks (*3*) also called knowledge network described by Newman. The most important examples of information network are the citation network (*8, 9*) and the World Wide Web (*10*). However, these networks neither focus on the knowledge themselves, nor regard the links between vertices as the intensive states. Here, we propose a *weighted graph* (*11, 12*) by assigning weights to the interactions, named knowledge network which is generated from the reader's borrowing records in library. This knowledge network aims to gain a better understanding of the relationship between knowledge and reveal the structure of the knowledge system.

**Data:**

A significant portion of knowledge network is reconstructed from the library's borrowing lists of graduate students enrolling in Nanjing University in 2003. The number of graduate students is 2853. The data set covers their all study time until graduating, but excludes some subjects as literature, history, philosophy, military, arts and education, which is lack of strong specialty of knowledge background (anyone can borrow and read these subjects if they interest). The data set mainly contains the following subjects: nature science, medicine, sociology, economy, management, psychology and a great part of industrial technology. Every book is classified by the Chinese Library Classification (CLC) and has a CLC number. For example, the CLC number of 'The Algebraic Eigenvalue Problem' written by Wilkinson, J.H. is O151.21, suggests this book is belonged to the sub-subject of 'Matrix theory' under the subject of mathematics. Such a detailed sub-subject is defined as *knowledge* in this paper, represented by a vertex in the graph. Of course, there may be some books with different names under a CLC number, but we regard them as the same *knowledge*. Two kinds of knowledge are connected with an undirected link if someone had borrowed them both. We consider these connections denoting the natural relationship between two kinds of knowledge, because graduate students have a strong reading tendency. In another word, they borrow and read books which are relevant to their research topics, especially excluding the leisure reading such as history, arts, literature, and some other popular books.

Commonly, the graph is expressed by its adjacency matrix $a_{ij}$, whose elements take the value 1 or 0, according to whether there is a link between vertex $i$ and $j$. However, the data set mentioned above allows us to go beyond this topological representation. We obtain the weighted picture of knowledge network by assigning weights to the edges representing each connecting link. For the knowledge network, the intensity $w_{ij}$ of the interaction between the two kinds of knowledge $i$ and $j$ is defined as:

$$w_{ij} = \sum_l \delta_i^l \delta_j^l \qquad [1]$$

where the index $l$ runs over all reader's borrowing lists, $\delta_i^l$ is the times of knowledge $i$ showing up in the list $l$, and 0 if it does not appear. This definition seems to be objective of knowledge association, because the relationship of knowledge $i$ and $j$ is closer if they are borrowed more times by same readers. Thus, we construct a network representing the association and their intensity between knowledge, where $w_{ij} = w_{ji}$. It is sufficient detail for us to address the large-scale features of the knowledge system. The data set allows us to capture the relationship between the weights and network's topology and to gain the local cohesive structure. These explorations are beneficial to understand how the individuals of knowledge organized to a knowledge or information system. Here, we take the first step to demonstrate the role that the different class knowledge (high- or low- degree vertex) play in learning, especially in academic specialty community.

**Results:**

There is 3447 kinds of knowledge in data set totally. The knowledge network of the giant component comprises $N = 3,253$ vertices accounting for the knowledge and $E = 180,872$ (for the symmetric graph) edges denoting the association between knowledge. The average degree of the network $<k> = 2E/N = 112$, as well as the maximal degree is 1805. One thing should be noticed that the most connected vertex (vertex with maximal degree) is TP312 accounting for 'Algorithm and Programming' in CLC, which is distinctive because of its wildly using in most of fields. The maximal degree decreases to 865 after excluding the knowledge of TP312. In particular, the average shortest path length, measuring the average number of edges separating any two vertices in the network, shows that the value $<\zeta> = 2.489$ is very small compared with the network size $N$, while the average clustering coefficient $C = 0.703$. These results imply that the knowledge network has a small world phenomenon. Random graph with the same number of vertices, N and the same average degree, <k> is generated to compare with knowledge network. The average shortest path length of the random graph, 1.986, scales as $\ln N/\ln <k>$, which is slightly less than the real one, 2.489. Meanwhile, the average clustering coefficient of the random graph, 0.034, is much smaller than counterpart of real one. The above small average shortest path length and high clustering coefficient imply the small world property of knowledge network.

The degree distribution is found to be the superposition of two exponential distributions. It takes the form of $P(k) = c_1 * exp(-\lambda_1 * k) + c_2 * exp(-\lambda_2 * k)$, where $c_1$ and $c_2$ are constant and $\lambda_1 = 0.11$, $\lambda_2 = 0.05$. (see Fig.1). As mentioned above, knowledge network is therefore an example of a network with small average shortest path length and a heterogeneous degree distribution, showing a sharp exponential cut-off.

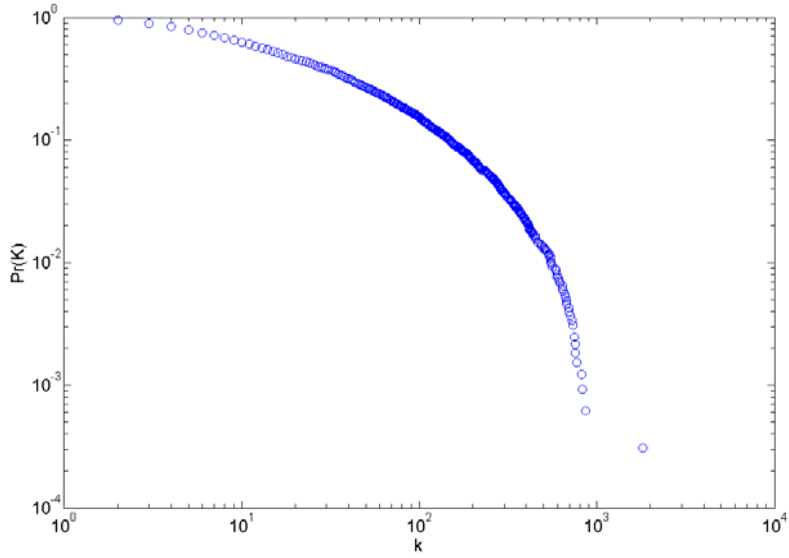

**Fig.1.** Degree distribution in the knowledge network. The degree $k$ accounts for how many knowledge connecting with the one knowledge. The distribution has a sharp exponential cut-off. It has the form $P(k) = c_1 * exp(-\lambda_1 * k) + c_2 * exp(-\lambda_2 * k)$.

**Topology:**

To investigate the information provided by the weighted graph, the appropriate quantities characterizing its statistical structure and organization should be distinguished. However, because the individual edge weight $w_{ij}$ between pairs of vertices can not provide a comprehensive picture of the complexity of weighed network (*13*), we use a more significant measure named vertex *strength* $s_i$ (*13,14*)

$$s_i = \sum_{j=1}^{N} a_{ij} w_{ij} \qquad [2]$$

This quantity measures the strength of vertices according to total weight of the connections, is a physical measure of centrality (importance) of vertices in the graph (*13*). In the case of knowledge network, the vertex strength accounts for the overall borrowing times of this knowledge, excluding the knowledge shown on the lists of those readers who have just borrowed one.

The knowledge network has a right-skewed strength distribution with an exponential cut-off (see Fig.2), indicating that though most knowledge are borrowed few times, a small minority are borrowed a great lot times. The tail is approximated by an exponential distribution $P(s) \sim e^{-\beta s}$, where $\beta=0.05$, exhibiting a similarity with the degree distribution $P(k)$. Such a rapid decay is maybe rooted in the fact that some little knowledge such as "Algorithm and Programming" will be borrowed by many readers because they are used in almost every field, but the vast remained knowledge are just borrowed in a small range of readers, for instance, "Quantum chromodynamics" is nearly spreading in a very small community of theoretical physics. The degree and strength distributions' cut-off property suggests the characteristics of statistical fluctuations varying with the network size. Namely, the network we construct is very heterogeneous.

One of the important issues in the research of complex network is to identify the most central vertices in system (*15*). Degree and betweenness centrality (*15,16,17*) are two common and natural quantities used to measure the centrality in topological networks. However, they neglect whether the relationship between the vertices is strong or weak, so they can not capture the characteristics of physical intensity of connection accurately in the real systems. Therefore, when we consider about the effect of strength of interaction, the definition of *strength* becomes more natural and proper in the field of weighted networks. It gives attention to both the number of connection and the intensity between them.

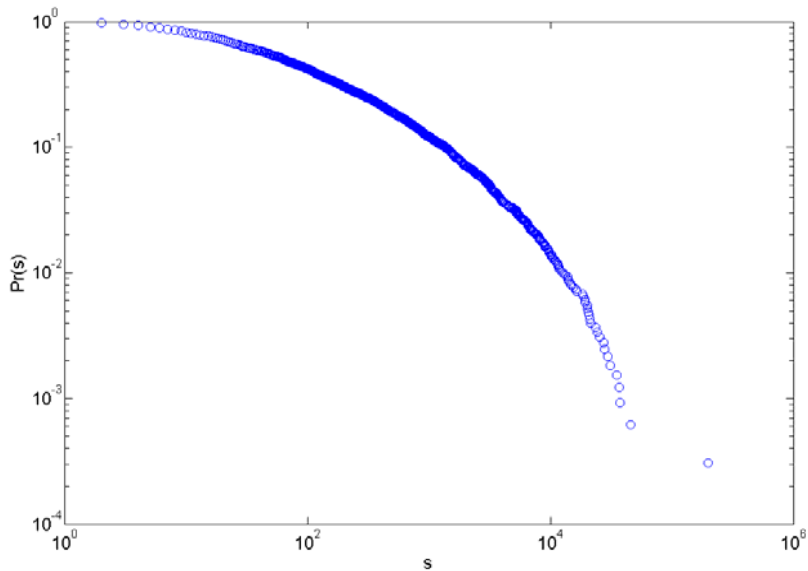

**Fig. 2.** Strength distribution in the knowledge network. The strength $s_i$ corresponds to the overall borrowing times of the knowledge $i$. The distribution can be approximated by a exponential distribution $P(s) \sim e^{-\beta s}$, where $\beta = 0.05$.

To uncover the relationship between the vertices' strength and graph's topology, we investigate the dependence of $s_i$ on $k_i$. We find that the average strength $s(k)$ of vertices with degree $k$ increases with the degree as a power law function:

$$s(k) \sim k^Y \quad [3]$$

where $Y=2.4\pm0.1$. In the absence of the correlations between the weight of edges and the degree of vertices, we can approximate $w_{ij} = <w> =(2E)^{-1}\sum_{i,j} a_{ij}w_{ij}$, where $<w>$ is the average weight of the graph. Then we have $s_i=<w>\sum_j a_{ij}=<w>k_i$ from Eq. **2**. So, the strength of a vertex is simply proportional to its degree, where $Y=1$ of Eq **3**. For simulating the above uncorrelated situation, a random picture of the real weighted network was generated through random reassigning of the actual weights on the existing topology of the graph. The behavior of the real data is very different with the random one (see Fig.3.). The comparison between the two curves indicates the strength of vertices grows faster than their degree. Thus the strength of vertices with high degree have a progressive tendency to be higher than the one in the graph with random assignment of weights, while the strength of vertices connects with little is prone to have a value lower than the one in the random picture. In the case of knowledge network, it implies if knowledge is shared by other more knowledge, it will be borrowed more times. Thus, there is a strong correlation between the weight and the topology presented by this tendency.

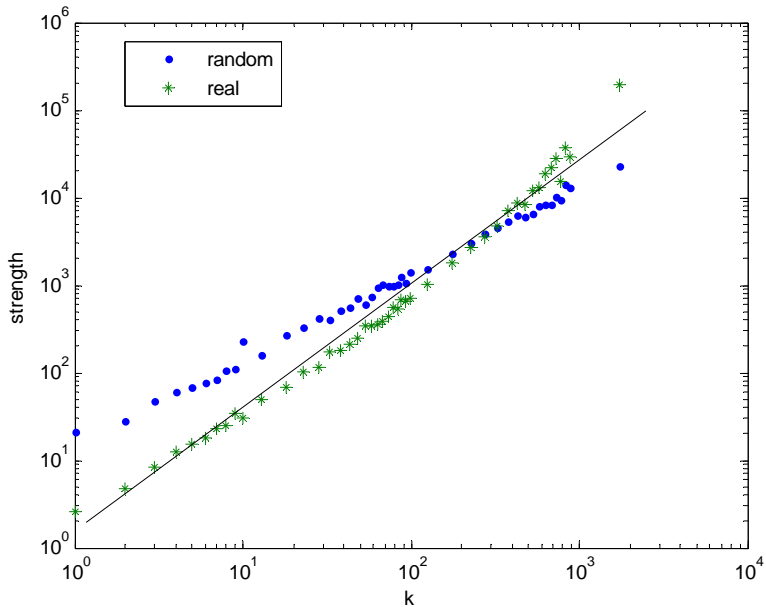

**Fig.3** Average strength $s(k)$ as function of the degree $k$ of vertices. The real data follows a power law behavior with $Y=2.4\pm0.1$, comparing with the random picture, where $Y=1$. The difference

between two curves suggesting anomalous correlation between the overall borrowing times of one knowledge and its number of neighbors.

**Structural Organization:**

The architecture imposed by the structural organization can not be fully characterized by the statistical properties as distribution *P(k)* and *P(s)*. A set of quantities are studied in order to uncover the structure hidden in the correlations between the vertices. The first one is the clustering coefficient $c_i$, which measures the local cohesiveness of topological structure (*18*). Thus the average clustering coefficient *C* denotes the global density of interconnected vertex triplets at the statistical level of the network. The function of average clustering coefficient as class degree *C(k)* could be used to unearth the hierarchical organization of the network. Some works observed a highly nontrivial behavior with rapidly decreasing of curves of *C(k)* in real networks, which suggests that low-degree nodes belongs to a highly cohesive, well interconnected clusters (high clustering coefficient), the hubs have a low probability to connect each other (small clustering coefficient). Another important quantity is the average nearest-neighbor degrees (19), $k_{nn}(k)$, as the function of classes *k*, measuring the correlations of the degree of neighboring vertices (*20*). In the absence of correlation, the function of $k_{nn}(k)$ does not depend on the *k*, i.e., $k_{nn}(k)$=constant (19). On the contrast, in the presence of correlation, the curves of $k_{nn}(k)$ can be classified to two classes. If vertices with large degree have a tendency to connect with the large degree vertices, $k_{nn}(k)$ is the increasing function of *k*, while $k_{nn}(k)$ is a decreasing curve, when well connected vertices have a majority of low degree neighbors. These two behaviors are referred as *assortative mixing* and *disassortative mixing* (*21*), respectively.

These two quantities above are defined under the topological background, but can not provide a deep signature of importance of weights in the forming of topological structure. Thus, alone with them, we employ other two measures here. The first one is the weighted clustering coefficient (13,14):

$$c_i^w = \frac{1}{s_i(k_i-1)} \sum_{j,h} \frac{(w_{ij}+w_{ih})}{2} a_{ij} a_{ih} a_{jh} \qquad [4]$$

The local cohesiveness structure accounting for the clustered structure originated from amount of interaction magnitude found on the local triplets can be measured by this quantity. We define $C^w$ and $C^w(k)$ as the average weighted clustering coefficient of the weighed graph and the weighted clustering coefficient averaged over all vertices with degree $k$ respectively. Comparing the $C^w$ and $C$ is a good way to reveal the global correlation between the weights and topology. There are two occurrences in the real weighted graph. The graph will result $C^w > C$ if the triangles are more likely formed by the edges with large weights. In the opposite case, if the topological clustering is generated by edges with low weights, the graph will bring on $C^w < C$ (13,14). $C^w(k)$ and $C(k)$ have the same behavior with regard to the degree classes $k$.

Along with the weighted clustering coefficient, the *weighted average nearest-neighbors degree* defined as (13,14):

$$k_{nn,i}^w = \frac{1}{s_i} \sum_{j=1}^{N} a_{ij} w_{ij} k_j \qquad [5]$$

was employed here. This quantity can be used to reveal the weighted assortative or disassortative properties according to the intensity of the actual interactions, which means the vertices with specific weights tend to point to high- or low- degree neighbors. $k_{nn,i}^w > k_{nn,i}$ implies the edges with larger weights connect with neighbors with larger degree, and $k_{nn,i}^w < k_{nn,i}$ indicates the opposite case (*13,14*). Noticeably, the topological definitions of $c_i$ and $k_{nn,i}$ are the simplicity forms of the weighted definitions of $c_i^w$ and $k_{nn,i}^w$. We alter the non-zero values of the adjacency matrix of weighted graph to 1, and remain zero values, in order to change the weighted graph to the corresponding topological one. In this way, it is easy to see that $c_i^w = (s_i(k_i-1))^{-1} \sum_{j,h} a_{ij} a_{ih} a_{jh} *$ $(w_{ij}+w_{ih})/2 = (k_i(k_i-1))^{-1} \sum_{j,h} a_{ij} a_{ih} a_{jh} = c_i$, and $k_{nn,i}^w = s_i^{-1} \sum a_{ij} w_{ij} k_{ij} = k_i^{-1} \sum a_{ij} k_{ij} = k_{nn,i}$ respectively, because an adjacency matrix of unweighted networks just have 1 and 0 elements only, and $s_i=k_i$, $w_{ij} = 1$, and $(w_{ij}+w_{ih})/2 = 1$.

Therefore, the regular topological quantities and the relevant weighted ones are obtained both. The $C(k)$ exhibits a continuously decaying spectrum (see Fig. 4) suggesting that hubs have a much lower probability to belong to the well interconnected clusters than low-degree vertices. This behavior roots in the phenomena that well connected knowledge often associate with the different knowledge communities, which leads to the lower clustering coefficient, while the knowledge relating to few vertices usually form a well interconnected community (high clustering). Thus, the role of the well connected knowledge is more likely to bridge small communities of clusters to an integrated network. Then, the knowledge network shows a disassortative behavior in the whole $k$ spectrum (see Fig. 5) agreed with the mixing pattern of technological and biological networks (*21*). This clearly exhibits the existence of nontrivial correlation properties for the knowledge network. It is unexpected somehow, because it is not as same as social network' mixing patterns such as scientist collaboration network (*13*). We find that $C^w / C \approx 1.11$, signaling that the interconnected triplets are more likely to be formed by the edges with high weights. The same thing happens in the comparing of $C^w(k)$ and $C(k)$. For k > 8, the weighted clustering coefficient is always larger than the topological one. It is means that relationship of high intensity tends to cluster together. Readers are likely to learn a group of high-degree and well interconnected knowledge more often. It is very similar with the scientist collaboration network (SCN), in which the $C^w(k)$ is always larger than the $C(k)$ when *k* exceed the threshold (The threshold of *k* is 10 in the SCN). Finally, it is interesting to comparing the function of $k_{nn}^{w}(k)$ and $k_{nn}(k)$. While $k < 5$, $k_{nn}(k)$ approaches a constant value. It implies vertices with different degrees have a similar neighborhood, meaning it is an uncorrelated structure in the range of $1 \leq k \leq 5$. This behavior partly likes the the world-wide airport network (WAN), in which the $k_{nn}(k)$ is a constant when $k>10$. Thus, as the knowledge network, the WAN also shows the absencing of topological correlations when k is smaller or larger than a threshold. The value of $k_{nn}^{w}(k)$ is almost a constant, contrasting with the decaying spectrum of a long range of $k_{nn}(k)$, but the weighted one is bigger than the topological one in the all range. It provides a scenario where stronger relationship between knowledge tends to point to higher-degree knowledge.

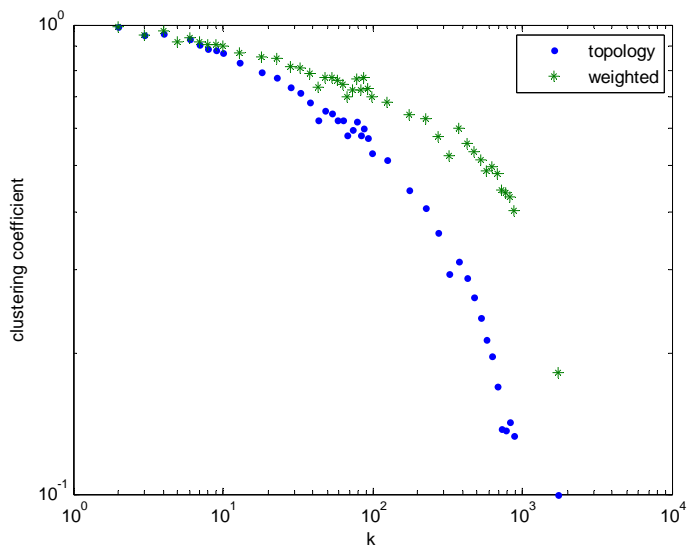

Fig. 4. Topological and weighted clustering coefficient. The weighted clustering gets apart from the topological one in the range of $k > 8$.

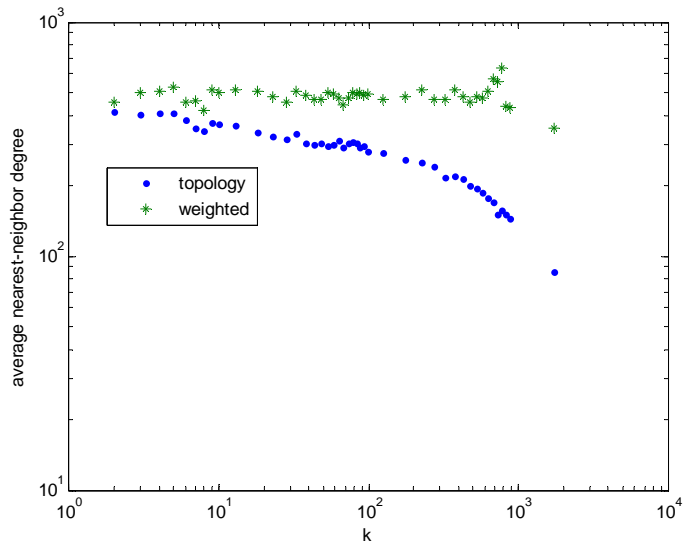

Fig.5 Topological and weighted average nearest-neighbors degree. $k_{nn}(k)$ is almost a constant. $k_{nn}^{w}(k)$ is a constant when $k \leqslant 5$, accounting for the absence of topological correlation, then approach to a continuous decaying spectrum.

**Conclusions**

In this paper we have studied knowledge network in which the nodes present knowledge classified by CLC, and a tie between two nodes represents distinct individuals who borrowed them both. Some statistics for network have been calculated, including numbers of knowledge per reader and times of borrowing per knowledge. The degree and the strength both exhibit a characteristic of statistical fluctuations with a continuous decaying spectrum, and the distributions of these quantities roughly follow an exponential form. We note that our network have the small world phenomena through the comparing between the real network and the random network.

We also demonstrate the correlation between weights and topology in a real system. The topological local connectivity is illustrated by varying degree classes in the structural organization. In particular, we explore the importance of weights in the process of forming local connectivity structure, which is edged with larger weight tend to form well connected clusters. Our analysis indicates that the hub may pay the role of connecting small clusters to the whole network. Our network shows a disassortiative behavior, which means high degree knowledge often have low degree neighbor. Lastly, in some range of degree k, the network is absence of topological correlation, which is agreed with some other network as world-wide airport network.

**Acknowledgment:** We thank the library of Nanjing University for providing the data of borrowing lists of readers to us. We also thank H.Q. Zhu, Y. Xie, X.K. Tian and R.L. Wang for many beneficial discussions. We are gratefully to the funding of Library of Nanjing University.